\documentclass{article}
\usepackage{spconf,graphicx,url}
\usepackage{amsmath,amssymb}
\usepackage{CJKutf8}
\usepackage{hyperref}
\usepackage{xspace}
\usepackage{ifthen}
\usepackage{booktabs}
\usepackage{subcaption}
\usepackage{hwemoji}
\usepackage{tikz}
\usepackage{enumitem}
\newcommand{\Fig}[1]{Figure~\ref{fig:#1}} 
\newcommand{\Table}[1]{Table~\ref{tbl:#1}} 
\newcommand{\Sec}[1]{Section~\ref{sec:#1}} 

\newcommand{\Section}[1]{
    \vspace{-1.8mm}
    \section{#1}
    \vspace{-1.5mm}
}
\newcommand{\Subsection}[1]{
    \subsection{#1}
}

\newcommand{\Subsubsection}[1]{
    \vspace{-2mm}
    \subsubsection{#1}
    \vspace{-1mm}
}

\newcommand{\zj}{\,$\downarrow$}
\newcommand{\zk}{\,$\uparrow$}

\newboolean{blind}
\setboolean{blind}{true}

\ifthenelse{\boolean{blind}}{
    \newcommand{\modelname}{THE T05 SYSTEM FOR THE VOICEMOS CHALLENGE 2024\xspace}
}{
    \newcommand{\modelname}{UTMOSv2\xspace}
}

\renewcommand{\Vec}[1]{\textrm{\boldmath $#1$}} 
\newcommand{\x}{ \Vec{x} } 
\newcommand{\w}{ \Vec{w} } 
\newcommand{\h}{ \Vec{h} } 
\newcommand{\e}{ \Vec{e} } 
\newcommand{\y}{ \Vec{y} } 

\title{\modelname: \\ Transfer Learning from Deep Image Classifier to Naturalness MOS Prediction of High-Quality Synthetic Speech}
\name{Kaito Baba, Wataru Nakata, Yuki Saito, Hiroshi Saruwatari}
\address{The University of Tokyo, Japan}

\begin{document}
\setlength{\dbltextfloatsep}{14pt} 
\ninept
\maketitle
\begin{abstract} 
We present our system (denoted as T05) for the VoiceMOS Challenge (VMC) 2024.
Our system was designed for the VMC 2024 Track 1, which focused on the accurate prediction of naturalness mean opinion score (MOS) for high-quality synthetic speech.
In addition to a pretrained self-supervised learning (SSL)-based speech feature extractor, our system incorporates a pretrained image feature extractor to capture the difference of synthetic speech observed in speech spectrograms. 
We first separately train two MOS predictors that use either of an SSL-based or spectrogram-based feature.
Then, we fine-tune the two predictors for better MOS prediction using the fusion of two extracted features.
In the VMC 2024 Track 1, our T05 system achieved first place in 7 out of 16 evaluation metrics and second place in the remaining 9 metrics, with a significant difference compared to those ranked third and below.
We also report the results of our ablation study to investigate essential factors of our system.
\end{abstract}
\vspace{-3pt}
\Section{Introduction}
\vspace{-3pt}
\label{sec:intro}
Automatic quality assessment of synthetic speech is an emerging research topic in the text-to-speech (TTS) and voice conversion (VC) research fields~\cite{mosnet,9746395}.
It is a promising technology for further development of TTS and VC because it can reduce the cost of human-based subjective evaluations on synthetic speech, such as a mean opinion score (MOS) test.
In fact, UTMOS~\cite{saeki22c_interspeech}, an open-sourced MOS prediction system, was introduced as an alternative way to compare the performances of TTS systems submitted to the Interspeech 2024 Speech Processing Using Discrete Speech Unit Challenge~\cite{chang24discrete_challenge}. 
Therefore, a MOS prediction system specialized for high-quality synthetic speech is valuable for a unified comparison of state-of-the-art deep neural network (DNN)-based TTS/VC systems~\cite{ju24naturalspeech3}.

The range-equalizing bias in MOS tests~\cite{cooper23interspeech} is one challenge to be addressed for achieving this goal. 
That is, listeners in a MOS test tend to use the entire range of choices on the rating scale (e.g., from one to five), regardless of the absolute quality of the samples used in the MOS test. 
For example, a medium-quality TTS/VC system in one MOS test may achieve relatively low MOS in another test excluding worse-performing systems from the comparison (i.e., zoomed-in MOS test).
Therefore, MOS prediction systems built without considering the range-equalizing bias may underestimate high-quality synthetic speech or overestimate low-quality synthetic speech.

In this paper, we present our MOS prediction system specialized for high-quality synthetic speech, which is designed for the VoiceMOS Challenge (VMC) 2024~\cite{huang24slt} Track 1, the task of predicting zoomed-in MOS test results.
Our system adopts some techniques that can improve the MOS prediction performance in the VMC 2022 and 2023~\cite{huang22f_interspeech, vmc2023}: using self-supervised learning (SSL)-based speech features~\cite{saeki22c_interspeech} and fusing multiple speech features~\cite{qi23le_ssl_mos}.
We also investigate the effectiveness of using EfficientNetV2~\cite{tan21efficientnetv2}, i.e., DNN-based {\it image} feature extractor, for capturing the difference of synthetic speech observed in speech spectrograms accurately.
In our two-stage fine-tuning strategy, we first separately train two MOS predictors that use either of an SSL-based or spectrogram-based feature.
Then, we fine-tune the two predictors for better MOS prediction using the fusion of two extracted features.
In the VMC 2024 Track 1, our T05 system achieved first place in 7 out of 16 evaluation metrics and second place in the remaining 9 metrics, with a significant difference compared to those ranked third and below.
We also report the results of our ablation study to investigate essential factors of our systems. 
The result demonstrates that fusing the two features improves the correlation-based evaluation metrics.
It also indicates that using a large-scale MOS dataset consisting of solely neural TTS samples or an actual zoomed-in MOS dataset for the training enhances the MOS prediction performance.
The code and the demo for our system are available online\footnote{\label{urls}
Code: \url{https://github.com/sarulab-speech/UTMOSv2}\\
\hspace*{5.75mm}Demo: \url{https://huggingface.co/spaces/sarulab-speech/UTMOSv2}}. 
\Section{The VMC 2024 Track 1}
\label{sec:voicemos2024}

The VMC 2024~\cite{huang24slt} consists of three tracks, where our T05 system is designed for the Track 1.
In this track, the organizers collected the results of zoomed-in MOS tests, where they compared speech synthesis systems that achieved high MOS from the BVCC dataset~\cite{cooper21_ssw}.
The organizer conducted three MOS tests with the zoom-in rates of 50\%, 25\%, and 12\%, representing the number of systems covered in the test compared to the original BVCC dataset.
No official training data considering these ``zoomed-in'' situations were provided by the organizers, and thus participants were required to build their MOS prediction systems with publicly available MOS datasets.
After the track finished, the organizers disclosed that the validation set consisted of the results of 50\% zoomed-in MOS test, while the evaluation set consisted of both 25\% and 12\% zoomed-in MOS tests.
The evaluation metrics included mean squared error (MSE), linear correlation coefficient (LCC), Spearman's rank correlation coefficient (SRCC), and Kendall's rank correlation coefficient (KTAU) at both the utterance and system levels. 
\setlength{\abovedisplayskip}{4pt}
\setlength{\belowdisplayskip}{4pt}

\Section{Our Submitted System (UTMOSv2)}
\label{sec:submitted-system}

\Subsection{Basic Architecture}

\begin{figure*}[tb]
    \centering
    \begin{minipage}[b]{.55\textwidth}
        \centering
        \includegraphics[width=98mm, trim=280 310 250 80, clip]{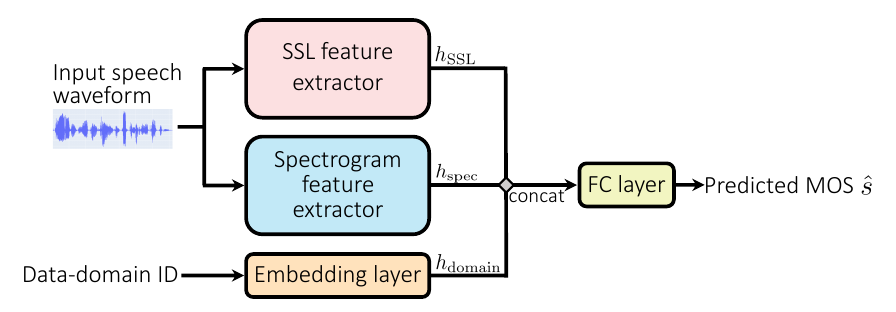}
        \subcaption{Basic architecture.}
        \label{fig:basic_architecture}
    \end{minipage}
    \begin{minipage}[b]{.44\textwidth}
        \centering
        \includegraphics[width=72mm, trim=250 40 360 350, clip]{figure/architecture2.pdf}
        \subcaption{Spectrogram feature extractor.}
        \label{fig:spec_architecture}
    \end{minipage}
    \caption{The basic model architectures in the proposed system (UTMOSv2). Our system leverages SSL features $\h_\text{SSL}$ and spectrogram features $\h_\text{spec}$. Additionally, the data domain embedding $\h_\text{domain}$ is obtained from the data-domain ID which is unique to each dataset used in the training. Finally, these three features are concatenated to predict the MOS of the input speech.}
\end{figure*}

Our T05 system (UTMOSv2) leverages the combination of spectrogram features extracted by a pretrained image feature extractor and speech features obtained from pretrained speech SSL models (i.e., SSL feature). 
\Fig{basic_architecture} illustrates the basic model architecture.

\Subsubsection{Spectrogram Feature Extractor}
\label{sec:spec-feat}

The field of computer vision using deep learning has significantly advanced in recent years, and applying DNN-based models (i.e., deep image classifiers) to spectrograms has demonstrated promising results in some audio/speech processing tasks ~\cite{8928f2fa01164d148735e057879ea6f1,chung18b_interspeech,8272618}. 
Our system thus leverages the features extracted from spectrograms using a convolutional neural network (CNN) pretrained on a large image dataset. 
The architecture of our spectrogram feature extractor is shown in \Fig{spec_architecture}.

In spectrogram feature extraction, the input speech waveform is first transformed into multiple mel-spectrograms. Each mel-spectrogram is extracted with different short-term Fourier transform (STFT) settings, which aims to mitigate the problem of the trade-off between frequency resolution and time resolution determined by the window size~\cite{8928f2fa01164d148735e057879ea6f1}. 
Let $\x = [\x_1^\top, \ldots, \x_K^\top]^\top$ be $K$ speech frames, where $\x_k$ denotes the $k$th frame consisting of $L$ samples. These frames are randomly extracted from the input speech waveform. Multiple mel-spectrogram transformations with $N$ different window sizes $(w^{(1)},\cdots, w^{(N)}), w^{(n)} \in \mathbb{N}$, $\mathrm{MelSpec}_{w^{(n)}}(\cdot)$, are applied to each extracted audio frame:
\[
   \y_{k}^{(n)} = \mathrm{MelSpec}_{w^{(n)}}(\x_k),
\]
where $\y_{k}^{(n)}$ denotes the $n$th mel-spectrogram extracted from the $k$th speech frame using the window size $w^{(n)}$.

These mel-spectrograms are then regarded as images rather than speech parameter sequences and fed into CNNs pretrained on ImageNet~\cite{imagenet}, following previous work~\cite{8272618}. The shape of mel-spectrogram image is fixed as $(F, F)$, where $F$ represents the number of mel-bands, regardless of the window size setting. 
Multiple CNNs are prepared, where each network receives a spectrogram with a different window size $w^{(n)}$ as input, extracting an image feature as follows:
\[
   \h_{k}^{(n)} =  \mathrm{CNN}_{w^{(n)}}(\y_{k}^{(n)}).
\]
The features obtained from $\y_{k}^{(n)}$ through the CNN for each window width setting, i.e., $(\h_{k}^{(1)}, \ldots, \h_{k}^{(N)})$, are aggregated using a weighted sum $\tilde{\h}_k = \sum_{n=1}^{N} w_{\mathrm{spec},n} \h_{k}^{(n)}$.
The trainable weight parameter vector $\w_{\text{spec}} \in \mathbb{R}^{N}$ is initialized such that $\sum_{n=1}^N w_{\text{spec},n} = 1$. As a result, the aggregated feature $\tilde{\h}_k$ has the dimension $\mathbb{R}^{c\times f\times t}$, where $c, f$, and $t$ denote the number of features, the height of feature maps and the width of the feature maps obtained through CNNs, respectively. These aggregated features with different $k$ are then concatenated across several frames in the $t$ dimension and subsequently pooled in both $t$ and $f$ dimensions.
A combination of average and max pooling is used in the time direction; a combination of attention~\cite{NIPS2017_3f5ee243} and max pooling was employed in the frequency direction.
The final output of our spectrogram feature extractor is hereinafter denoted as $\h_{\text{spec}}$.

\Subsubsection{SSL Feature Extractor}
\label{sec:ssl-feat}

Following previous studies on automatic MOS prediction~\cite{9746395,saeki22c_interspeech}, we utilize a pretrained SSL model to extract speech features from an input waveform. The raw waveform is first fed into the SSL model to extract hidden states from the each layer of the Transformer encoder $(\e_1, \e_2,\cdots, \e_M)$. Then, the hidden states are aggregated using a weighted sum $\tilde{\e} = \sum_{m=1}^{M} w_{\text{SSL},m} \e_{m}$, where $M$ denotes the number of Transformer encoder layers.
The trainable weight parameter vector $\w_{\text{SSL}} \in \mathbb{R}^{M}$ is initialized such that $\sum_{m=1}^M w_{\text{SSL},m} = 1$.
Finally, unlike in previous studies~\cite{9746395,saeki22c_interspeech}, combination of 
attention~\cite{NIPS2017_3f5ee243} and max pooling along the sequence dimension are applied to the aggregated hidden state vectors for each time step. 
The final output of our SSL feature extractor is hereinafter denoted as $\h_{\text{SSL}}$.

\Subsubsection{Data-domain Encoding}
\label{sec:data-enc}

Following the UTMOS system~\cite{saeki22c_interspeech}, we build our MOS prediction system using multiple MOS datasets for the model training with the data-domain encoding (i.e., conditioning the system on the dataset ID).
This aims to address the biases in different MOS tests, possibly including the range-equalizing bias~\cite{cooper23interspeech}.
For the data-domain encoding, simple look-up embedding table is used for converting discrete dataset ID to continuous data-domain embedding $\h_{\text{domain}}$.

Note that this data-domain encoding cannot define IDs for unseen MOS datasets and thus does not necessarily work properly for the out-of-domain prediction.
One can deal with this issue by, for example, predicting MOS for some seen data-domains and taking the average of multiple predicted scores~\cite{saeki22c_interspeech}.
Because the primal focus of this paper is the range-equalizing bias, we thoroughly investigate the domain gap between the training and test datasets in our ablation study (\Sec{ablation_data}).

\Subsubsection{Fusion of Spectrogram Features and SSL Features}
A simple fully connected layer is prepared and trained for predicting the MOS of input speech using the fusion of extracted spectrogram and SSL features denoted as $\h_\text{spec}$ and $\h_\text{SSL}$, respectively.
The input is the concatenation of these two features and the data-domain embedding along the feature dimension:
\begin{equation}\label{eq:fusion}
   \hat{s} = \operatorname{FC} \left( \operatorname{Concat}(\h_\text{spec}, \h_\text{SSL}, \h_\text{domain})\right) ,
\end{equation}
where $\operatorname{FC}(\cdot)$ and $\operatorname{Concat}(\cdot)$ denote a fully connected layer and feature concatenation, respectively.

\Subsection{Additional Data Collection}
\label{sec:ext-data}
\begin{table}[t]
    \centering
    \caption{Specifications of dataset used in the training. ``BC'' and \# mean ``Blizzard Challenge'' and ``number of,'' respectively.}
    \label{tbl:dataset_information}
    \scalebox{0.95}{
    \begin{tabular}{l|r|r|r|r}\toprule
         Dataset name& \# listeners  & \# systems & \# sentences & \# ratings \\\midrule
         BC2008 & 229 & 7 & 80 &  16,987 \\
         BC2009 & 129 & 19 & 141 &  21,332 \\
         BC2010 EH1 & 177 & 18 & 36 &  5,863 \\
         BC2010 EH2 & 179 & 18 & 36 &  6,070 \\
         BC2010 ES1 & 73 & 8 & 16 &  1,152 \\
         BC2010 ES3 & 84 & 8 & 16 &  1,250 \\
         BC2011 & 236 & 13 & 39 &  9,328 \\
         BVCC & 304 & 187 & 7106 & 56,848 \\
         SOMOS & 987 & 201 & 2000 &  359,100 \\
         sarulab-data & 304 & 95 & 3610 & 28,880 \\\bottomrule
    \end{tabular}
    }
    
\end{table}

As there are no official training sets provided by the organizers, we collected training data from the publicly available MOS test results.
The collected data consisted of BVCC~\cite{cooper21_ssw}, Blizzard Challenge (BC) 2008~\cite{blizzard2008}, 2009~\cite{blizzard2009}, 2010~\cite{blizzard2010}, 2011~\cite{King2011TheBC} SOMOS~\cite{maniati22_interspeech}, and zoomed-in BVCC dataset that is publicly available (sarulab-data)\footnote{\url{https://github.com/sarulab-speech/VMC2024-sarulab-data}}.
The specification of datasets are shown in \Table{dataset_information}.

For the dataset derived from BC, we only used subjective evaluation results for the english utterances. For BC2008, We excluded listeners which are marked with EUS as their scores were not in 5-point scale. 
For BC2010, We used results for task EH1, EH2, ES1 and ES3. ES2 was excluded as naturalness of synthetic speech was not considered in this task.


\Subsection{Loss Function}
For the loss function used in the training, we adopt the combination of a contrastive loss~\cite{saeki22c_interspeech} and mean squared error (MSE) loss.
Specifically, the contrastive loss is formulated as
\begin{equation}\label{eq:con}
\mathcal{L}_{\text{con}}(s, \hat{s}) = \sum_{i\neq j}\max(0, |(s_i - s_j) - (\hat{s}_i - \hat{s}_j)| - \alpha),
\end{equation}
where $s$ and $\hat{s}$ denote the target MOS and predicted MOS, respectively.
The margin hyperparameter $\alpha > 0$ makes the trained model ignore small errors lower than this margin.
The final loss $\mathcal{L}$ is defined as follows:
\begin{equation}\label{eq:mse}
    \mathcal{L}(s, \hat{s}) = \lambda_{\text{con}} \mathcal{L}_{\text{con}}(s, \hat{s}) + \lambda_{\text{mse}} \mathcal{L}_{\text{mse}}(s, \hat{s} ),
\end{equation}
where $\lambda_{\text{con}}$ and $\lambda_{\text{mse}}$ are hyperparameters that control the weights of the contrastive and MSE loss functions, respectively.

\Subsection{Multi-Stage Learning}
\label{sec:two-stage-learning}

When fine-tuning a pretrained model, catastrophic forgetting can significantly worsen the performance of the model on learned domains~\cite{french99}. To mitigate this, we introduce multi-stage learning.

Since our proposed system is large and difficult to train the parameters of two feature extractors from scratch, we first train the two extractors separately.
Then, we fine-tune the pretrained weights from these individual models and train the parameters of the $\operatorname{FC}$ layer (Eq.~\ref{eq:fusion}) for the feature fusion. In summary, the training performs the following stages:
\begin{enumerate}[label=Stage \arabic*:]
    \setlength{\itemindent}{15pt} 
    \item The spectrogram and SSL feature extractors are trained separately. Specifically, an $\operatorname{FC}$ layer, which takes the concatenated features of data-domain embedding $\h_\text{domain}$ and either of $\h_\text{spec}$ or $\h_\text{SSL}$ and predicts MOS, is trained jointly with the extractor.
    \item The weights of the two extractors are frozen and only the feature fusion layer (Eq.~\ref{eq:fusion}) along with a new data-domain embedding layer is trained.
    \item All parameters of the models in our system are fine-tuned with a small learning rate.    
\end{enumerate}
Our SSL feature extractor is also pretrained with two-stage training following similar stages described above.
That is, the model parameters of a backbone SSL model is first frozen and only the $\operatorname{FC}$ layer for the MOS prediction is trained.
Then, all parameters of this extractor including the SSL model are fine-tuned.
In contrast, our preliminary experiment showed that this two-stage pretraining for the spectrogram feature extractor did not bring significant improvement.
Therefore, we decided to train the entire model of this extractor, i.e., the pretrained CNNs and the $\operatorname{FC}$ layer for the MOS prediction.

Technically, in the comparative experiments in \Sec{experiment}, the spectrogram feature extractor was trained using data-domain embeddings on all datasets. Meanwhile, the system submitted for the VMC2024 Track 1 excluded the data-domain encoding and performed fine-tuning on sarulab-data after the training on BVCC. Apart from this aspect, the DNN architecture used in the comparative experiments in Section 4 and the submitted system is exactly the same. 
\Section{Experiments}
\label{sec:experiment}

We conducted several experiments to validate the effectiveness of our T05 system. Specifically, we performed ablation studies on the fusion of spectrogram and SSL features, multi-stage learning, and datasets.

\Subsection{Common Experimental Conditions}

We used EfficientNetV2~\cite{Tan2021EfficientNetV2SM} as the CNN for our spectrogram feature extractor. For the backbone SSL model, we used wav2vec2.0~\cite{NEURIPS2020_92d1e1eb} base\footnote{\url{https://huggingface.co/facebook/wav2vec2-base}} pretrained on LibriSpeech~\cite{panayotov2015librispeech}.
For the data-domain encoding, we used embedding with hidden size of 1.

For the loss function, we set the margin hyperparameter $\alpha=0.2$ (Eq.~(\ref{eq:con})) for all experiments. The weight coefficients for the contrastive and MSE loss, $\lambda_{\text{con}}$ and $\lambda_{\text{mse}}$, were set to of 0.2 and 0.7, respectively. These hyperparameters were decided based on our preliminary experiments.
For the optimizer, we used AdamW~\cite{loshchilov2018decoupled} with the weight decay coefficient of $1\times10^{-4}$. For learning rate scheduler, we decayed the learning rate with a cosine annealing~\cite{loshchilov2017sgdr}. 
The initial learning rate varied depending on the training stage. 
During training, we incorporated mixup~\cite{zhang2018mixup} for all training, which was shown to be effective in MOS prediction~\cite{wang-etal-2023-mospc}.

A five-fold cross-validation was performed, and the best model checkpoint was selected based on the average system-level SRCC calculated for each validation fold. The final prediction was obtained by averaging the predictions from each of the five folds. Additionally, during inference, we generated predictions five times by randomly selecting different frames of the input speech waveform and then averaged these predictions (i.e., test-time augmentation~\cite{shanmugam21tta}).

\begin{table*}[tb]
    \centering
    \setlength{\tabcolsep}{1mm} 
    \caption{Comparison of performance between our systems and the ``B01'' baseline. \textbf{Bold} and \underline{underlined} scores are the best and worst among our three systems, respectively. We also compare our systems and human-annotated MOS (``BVCC MOS'').}
    \label{tbl:comparison_of_models}
    \scalebox{0.88}{
        \begin{tabular}{l|cccc|cccc|cccc|cccc}
            \toprule
                                             &                                                            \multicolumn{8}{c}{Zoom-in rate: 25\%}                                                                      &                                                                     \multicolumn{8}{|c}{Zoom-in rate: 12\%} \\
                                             &                          \multicolumn{4}{c}{Utterance-level}                      &                                 \multicolumn{4}{|c}{System-level}                  &                                 \multicolumn{4}{|c}{Utterance-level}              &                          \multicolumn{4}{|c}{System-level}                        \\
                                             &             MSE\zj &             LCC\zk &            SRCC\zk &            KTAU\zk &              MSE\zj &             LCC\zk &            SRCC\zk &            KTAU\zk &             MSE\zj &             LCC\zk &            SRCC\zk &            KTAU\zk &             MSE\zj &             LCC\zk &            SRCC\zk &            KTAU\zk \\
            \midrule
            Ours                             &              0.690 &      \textbf{0.618}&      \textbf{0.613}&      \textbf{0.442}&               0.465 &      \textbf{0.922}&      \textbf{0.919}&      \textbf{0.752}&              0.459 &      \textbf{0.578}&      \textbf{0.579}&      \textbf{0.404}&              0.288 &      \textbf{0.840}&      \textbf{0.854}&      \textbf{0.650}\\
            w/o SSL                          &      \textbf{0.566}&   \underline{0.576}&   \underline{0.565}&   \underline{0.403}&      \textbf{0.353} &   \underline{0.889}&              0.909 &              0.740 &     \textbf{0.357} &   \underline{0.518}&   \underline{0.516}&   \underline{0.355}&     \textbf{0.188} &   \underline{0.762}&   \underline{0.770}&   \underline{0.570}\\
            w/o spec.                    &   \underline{0.937}&              0.602 &              0.603 &              0.432 &   \underline{0.700} &              0.910 &   \underline{0.909}&   \underline{0.731}&  \underline{0.673} &              0.530 &              0.529 &              0.364 &  \underline{0.497} &              0.793 &              0.793 &              0.570 \\
            \midrule
            B01                              &              1.154 &              0.508 &              0.509 &              0.358 &               0.998 &              0.750 &              0.745 &              0.539 &              0.741 &              0.422 &              0.417 &              0.285 &              0.589 &              0.608 &              0.609 &              0.444 \\
            UTMOS~\cite{saeki22c_interspeech}&              0.872 &              0.407 &              0.411 &              0.286 &               0.690 &              0.649 &              0.615 &              0.433 &              0.541 &              0.297 &              0.300 &              0.206 &              0.378 &              0.440 &              0.367 &              0.230 \\
            \midrule
            BVCC MOS                         &              0.717 &              0.377 &              0.358 &              0.256 &               0.413 &              0.728 &              0.679 &              0.495 &              0.481 &              0.322 &              0.316 &              0.225 &              0.223 &              0.691 &              0.702 &              0.467 \\
            \bottomrule
        \end{tabular}
    }
\end{table*}

\subsection{Evaluation Metrics}
The evaluation was performed on the test set with the zoom-in rate of 25\% and 12\%.
In both test sets, we used system level and utterance-level MSE, LCC, SRCC and KTAU as metrics, referring to the VMC2024 evaluation protocol.

\Subsection{VMC2024 Results of Our T05 System~\cite{huang24slt}}

In the Track1, both utterance-level and system-level metrics are calculated for 25\% and 12\% highest-rated systems, respectively. The official evaluation results show that our T05 system achieved the first place in 7 out of 16 metrics and ranked the second in the remaining 9 metrics, thereby securing either the first or second place in all metrics. Additionally, it is notable that there is a large margin in the performance to those ranked the third and below.

\Subsection{Ablation Study on Fusing Spectrogram/SSL Features}
\label{sec:comparison_of_fusion}

To evaluate the effectiveness of fusing spectrogram features and SSL features, we compared the prediction scores of the fused model with those obtained using only spectrogram or SSL features. 

\Subsubsection{Experimental Conditions}
In this ablation study, we compared the following systems:
\begin{itemize}
    \setlength{\itemindent}{-8pt}
    \itemsep -0.2mm
    \item {\bf Ours}: The proposed system using the feature fusion.
    \item {\bf Ours w/o SSL}: The proposed system using only the spectrogram feature extractor.
    \item {\bf Ours w/o spec.}: The proposed system using only the SSL feature extractor.
    \item {\bf B01:} SSL-MOS~\cite{9746395} trained on the original BVCC~\cite{cooper21_ssw} samples and labels. This system was considered as baseline system in the VMC 2024 track 1. 
    \item {\bf UTMOS~\cite{saeki22c_interspeech}:} The opensourced MOS prediction system.
\end{itemize}
``Ours w/o SSL'' was trained with a learning rate ranging from $1\times10^{-3}$ to $1\times10^{-7}$, a batch size of 10, and for 20 epochs. 
As explained in \Sec{two-stage-learning}, ``Ours w/o spec.'' was built with the two-stage training.
We first trained the $\operatorname{FC}$ layer and data-domain embedding for 20 epochs using the learning rate ranging from $1\times10^{-3}$ to $1\times10^{-7}$ and batch size of 32.
Then, we fine-tuned all model parameters for 5 epochs using the learning rate ranging from $3\times10^{-5}$ to $1\times10^{-9}$ and batch size of 32.
The system using the feature fusion, ``Ours,'' was built upon these two systems.
Specifically, we utilized these two feature extractors trained through ``Ours w/o spec.'' and ``Ours w/o SSL.''
The following $\operatorname{FC}$ layer, and data-domain embedding were randomly initialized.

The stage 2 training was performed for 8 epochs using a learning rate ranging from $1\times10^{-3}$ to $1\times10^{-5}$ and a batch size of 16.
The stage 3 training was iterated with 2 epochs using a learning rate ranging from $5\times10^{-5}$ to $1\times10^{-8}$ and a batch size of 8.

In this comparison, we used all datasets listed in Table~1 for the training and set the data-domain ID for the MOS prediction to ``BVCC'' in three our systems.

\begin{table*}[tb]
    \centering
    \setlength{\tabcolsep}{1mm} 
    \caption{
        Comparison of performance between our systems that did not employ the multi-stage learning process. \textbf{Bold values} are the best scores and \underline{underlined values} are the worst scores among each column.
    }
    \label{tbl:comparison_of_learning_step}
    \scalebox{0.85}{
        \begin{tabular}{l|cccc|cccc|cccc|cccc}
            \toprule
                                             &                                                            \multicolumn{8}{c}{Zoom-in rate: 25\%}                                                                      &                                                                     \multicolumn{8}{|c}{Zoom-in rate: 12\%} \\
                                             &                          \multicolumn{4}{c}{Utterance-level}                      &                                 \multicolumn{4}{|c}{System-level}                  &                                 \multicolumn{4}{|c}{Utterance-level}              &                          \multicolumn{4}{|c}{System-level}                        \\
                                             &             MSE\zj &             LCC\zk &            SRCC\zk &            KTAU\zk &              MSE\zj &             LCC\zk &            SRCC\zk &            KTAU\zk &             MSE\zj &             LCC\zk &            SRCC\zk &            KTAU\zk &             MSE\zj &             LCC\zk &            SRCC\zk &            KTAU\zk \\
            \midrule
            Ours                   &  \underline{0.690} & \textbf{0.618}&      \textbf{0.613}&      \textbf{0.442}& \underline{0.465} &      \textbf{0.922}&      \textbf{0.919}&      \textbf{0.752}&  \underline{0.459} &      \textbf{0.578}&      \textbf{0.579}&      \textbf{0.404}&    \underline{0.288} &      \textbf{0.840}&      \textbf{0.854}&      \textbf{0.650}\\
            w/o Stage 2    &  0.469 &                     0.531 &              0.555 &              0.394 &             0.209 &              0.900 &              0.911 &              0.744 &              0.342 &              0.436 &              0.505 &              0.350 &              0.108 &              0.787   &              0.816 &              0.602 \\
            w/o Stage 1\&2     &  \textbf{0.355}& \underline{0.480} &  \underline{0.482} &  \underline{0.336} &     \textbf{0.125}&  \underline{0.738} &  \underline{0.710} &  \underline{0.499} &      \textbf{0.293}&    \underline{0.421} & \underline{0.423} & \underline{0.289} &      \textbf{0.097}&   \underline{0.675}  &    \underline{0.672} & \underline{0.531} \\
            \bottomrule
        \end{tabular}
    }
\end{table*}

\Subsubsection{Results and Discussion}

The results are shown in \Table{comparison_of_models}.
For correlation-based metrics, we can see that all of our three systems consistently outperforms both two baseline models in all metrics. Furthermore, in correlation-based metrics, while there are little difference in scores between ``Ours w/o SSL'' and ``Ours w/o spec.,'' the fusion system, ``Ours,'' demonstrates a significant improvement in scores compared to these two systems. These results indicate that our systems are more effective for zoomed-in MOS prediction compared to the existing baseline systems, particularly in correlation-based metrics. It also suggests the effectiveness of fusing spectrogram and SSL features in these metrics.

One noteworthy observation is that ``Ours w/o SSL'' achieves the best MSE in all cases, but the worst in many cases in the correlation-based metrics. On the other hand, ``Ours w/o spec.'' scored the highest MSE, but outperforms ``Ours w/o SSL'' in many cases in the correlation-based metrics. From this perspective, we can infer that the spectrogram features derived from our image feature extractor are better at capturing fine differences in synthetic speech and predicting absolute MOS values, while SSL features are better at predicting rankings among multiple speech synthesis systems. In summary, these results suggest that the fusion of these features improves the prediction of absolute speech quality while further improving the correlation-based measures.


We also computed the evaluation metrics between the ground-truth MOS and human-annotated MOS (``BVCC MOS''), which was collected without considering the range-equalizing bias. The results from \Table{comparison_of_models} demonstrate that the bias actually exists and ``BVCC MOS'' is not well correlated with the ground-truth MOS. In contrast, our fusion system shows better scores than ``BVCC MOS'' in all metrics except for system-level MSE. Considering that the prediction is made with the data-domain embedding of BVCC, these results suggest that our system has demonstrated robust prediction of MOS for unseen listening test settings.

\Subsection{Comparison of Multi-Stage Learning}
To evaluate the effectiveness of the multi-stage learning described in \Sec{two-stage-learning}, we conducted a comparative experiment. 

\Subsubsection{Experimental Conditions}

In this experiment, we compared ``Ours'' in \Sec{comparison_of_fusion} with the following systems:
\begin{itemize}
    \setlength{\itemindent}{-8pt}
    \itemsep -0.2mm
    \item {\bf Ours w/o Stage 2}: The proposed system without performing the stage 2 training.
    \item {\bf Ours w/o Stage 1\&2}: The proposed system with performing only the stage 3 training.
\end{itemize}


The fine-tuning for ``Ours w/o Stage 2'' was performed for 20 epochs using a learning rate ranging from $1\times10^{-4}$ to $1\times10^{-7}$ and batch size of 8. 
The training for ``Ours w/o Stage 1\&2'' ran 20 epochs using a learning rate ranging from $1\times10^{-3}$ to $1\times10^{-7}$ and batch size of 8.
The training dataset and target domain-ID setting was the same as those used in \Sec{comparison_of_fusion}.

\Subsubsection{Results and Discussion}

The results are shown in \Table{comparison_of_learning_step}. 
As the number of multi-stage learning stages is reduced and the two feature extractors are no longer pre-trained for the MOS prediction task, the behavior of the learned models can be seen to approach ``Ours w/o SSL'' (i.e., lower MSE and lower correlation-based metrics). This may be desirable in situations where we want to accurately predict the absolute MOS, but not when we want to compare different speech synthesis systems. In summary, these results suggest that the proposed multi-stage learning is essential for boosting the ability of the SSL features to capture differences between multiple synthetic speech samples.

This might be because the SSL and spectrogram were combined and trained before being optimized individually. Due to the different learning speeds of SSL feature extractor and the spectrogram feature extractor, the fully connected layer might have resulted in a model that emphasizes one over the other. Specifically, in this case, the spectrogram features might have been given more importance, leading the system to resemble ``Ours w/o SSL.''


\Subsection{Investigation on Dataset}
\label{sec:ablation_data}

To investigate which datasets described in \Sec{ext-data} were effective for predicting the MOS for the zoomed-in target, i.e., newly obtained through listening tests of BVCC's top-performing systems, we conducted ablation studies on these datasets. 

\Subsubsection{Experimental Conditions}
For predicting MOS, we used ``Ours'' built with the almost same experimental setting as described in \Sec{comparison_of_fusion}.
Here, we changed the datasets for the training and the data-domain ID for the inference.
Specifically, we trained ``Ours'' with ``All datasets'' and that without \{BVCC, BC, SOMOS, sarulab-data\}.
This experiment enabled us to examine which dataset was essential for improving the MOS prediction performance in the zoomed-in test situation.
In addition, by examining the prediction results when changing the data-domain ID, we can verify which domain (i.e. dataset) was closer to the zoomed-in dataset used in the VMC2024 Track 1.
Note that only the mean values are presented in the results for the BC datasets, even though the data-domain ID was prepared for each BC dataset.

\Subsubsection{Results and Discussion}

The results are shown on \Table{ablation_study}.
In terms of the training datasets, ``All datasets'' achieves the best scores. However, in some cases the scores improve by excluding BVCC or BC from the training data.
In addition, excluding SOMOS or sarulab-data from the training data tends to degrade the MOS prediction performance significantly. These results suggest that when building MOS prediction systems to compare the performance of high-quality speech synthesis, it is crucial to exclude datasets that are likely to contain low-quality speech systems when training. They also indicate that using MOS datasets containing as many results as possible from evaluation of synthetic speech produced by state-of-the-art DNN-based speech synthesis. 

Focusing on the difference among data-domain for the MOS prediction, the MSE is the lowest for sarulab-data (i.e., the 50\% zoomed-in BVCC) and the highest for BVCC, which clearly shows the effect of range-equalizing bias~\cite{cooper23interspeech}. However, this tendency is not observed when comparing the correlation-based metrics. These results suggest that the negative effects caused by the range-equalizing bias are dominant in the prediction of the absolute MOS.


Additionally, when comparing the scores of correlation-based metrics between datasets with a 25\% zoomed-in rate and those with a 12\% zoomed-in rate, it can be observed that the scores are better for the 25\% zoomed-in rate datasets in almost all cases. This suggests that the quality of the speech data used for training was closer to that of the 25\% zoomed-in rate datasets.


\begin{table*}[tb]
    \centering
    \setlength{\tabcolsep}{1mm} 
    \caption{Results for ablation study regarding the training datasets and dataset domains. For example, the second-through-fifth columns list the MOS prediction performance for the VMC2024 Track 1 evaluation set using ``BVCC'' as the data-domain. Values in \textbf{bold face} shows the best result in each column and the \underline{underlined values} show the worst result in each columns. Only the mean values are presented for the BC datasets.}
    \label{tbl:ablation_study}
    \subcaption{Utterance-level results at 25\% zoomed-in rate.}
    \scalebox{0.85}{
        \begin{tabular}{l|cccc|cccc|cccc|cccc}
            \toprule
                                 &                \multicolumn{4}{c}{BVCC}                                           &                                     \multicolumn{4}{|c}{BC}                        &                                           \multicolumn{4}{|c}{SOMOS}                     &                                 \multicolumn{4}{|c}{sarulab-data}                        \\
            Training datasets    &             MSE\zj &             LCC\zk &            SRCC\zk &            KTAU\zk &                    MSE\zj &             LCC\zk &            SRCC\zk &            KTAU\zk &                    MSE\zj &             LCC\zk &            SRCC\zk &            KTAU\zk &                    MSE\zj &             LCC\zk &            SRCC\zk &            KTAU\zk \\
            \midrule
            All datasets         &              0.690 &     \textbf{0.618} &     \textbf{0.613} &     \textbf{0.442} &            \textbf{0.398} &              0.620 &              0.616 &              0.444 &            \textbf{0.326} &              0.617 &              0.612 &              0.441 &            \textbf{0.279} &              0.620 &              0.615 &              0.444 \\
            w/o BVCC             &                 -- &                 -- &                 -- &                 -- &         \underline{0.741} &     \textbf{0.668} &      \textbf{0.656} &    \textbf{0.480} &                     0.444 &     \textbf{0.668} &     \textbf{0.656} &     \textbf{0.480} &                     0.386 &     \textbf{0.667} &     \textbf{0.655} &     \textbf{0.479} \\
            w/o BC               &     \textbf{0.569} &              0.531 &              0.533 &              0.378 &                        -- &                 -- &                 -- &                 -- &                     0.414 &              0.528 &              0.527 &              0.374 &                     0.329 &              0.543 &              0.530 &              0.377 \\
            w/o SOMOS            &              0.683 &  \underline{0.417} &  \underline{0.411} &  \underline{0.286} &                     0.677 &  \underline{0.417} &   \underline{0.410} & \underline{0.286} &                        -- &                 -- &                 -- &                 -- &         \underline{0.678} &  \underline{0.416} &  \underline{0.408} &  \underline{0.285} \\
            w/o sarulab-data     &  \underline{0.733} &              0.473 &              0.470 &              0.329 &                     0.438 &              0.475 &              0.473 &              0.332 &          \underline{0.592} & \underline{0.474} &  \underline{0.470} &  \underline{0.330} &                        -- &                 -- &                 -- &                 -- \\

            \bottomrule
        \end{tabular}
    }
    \vspace{4mm}
    \subcaption{System-level results at 25\% zoomed-in rate.}
    \scalebox{0.85}{
        \begin{tabular}{l|cccc|cccc|cccc|cccc}
            \toprule
                                 &                                     \multicolumn{4}{c}{BVCC}                      &                                     \multicolumn{4}{|c}{BC}                        &                                           \multicolumn{4}{|c}{SOMOS}                     &                                 \multicolumn{4}{|c}{sarulab-data}                        \\
            Training datasets    &             MSE\zj &             LCC\zk &            SRCC\zk &            KTAU\zk &                    MSE\zj &             LCC\zk &            SRCC\zk &            KTAU\zk &                    MSE\zj &             LCC\zk &            SRCC\zk &            KTAU\zk &                    MSE\zj &             LCC\zk &            SRCC\zk &            KTAU\zk \\
            \midrule
            All datasets         &              0.465 &     \textbf{0.922} &     \textbf{0.919} &     \textbf{0.752} &            \textbf{0.167} &     \textbf{0.924} &     \textbf{0.919} &     \textbf{0.754} &            \textbf{0.092} &     \textbf{0.924} &              0.916 &              0.744 &            \textbf{0.044} &     \textbf{0.923} &     \textbf{0.921} &     \textbf{0.756} \\
            w/o BVCC             &                 -- &                 -- &                 -- &                 -- &         \underline{0.487} &              0.910 &              0.918 &              0.748 &                     0.194 &              0.911 &     \textbf{0.918} &      \textbf{0.748}&                     0.138 &              0.910 &              0.919 &              0.752 \\
            w/o BC         &           \textbf{0.284} &              0.912 &              0.916 &              0.746 &                        -- &                 -- &                 -- &                 -- &                     0.133 &              0.885 &              0.891 &              0.725 &                     0.051 &              0.899 &              0.911 &              0.744 \\
            w/o SOMOS            &              0.423 &  \underline{0.714} &  \underline{0.669} &  \underline{0.474} &                     0.415 &  \underline{0.714} &  \underline{0.664} &        \underline{0.472} &                        -- &                 -- &                 -- &                 -- &         \underline{0.416} &   \underline{0.708}&   \underline{0.655}&  \underline{0.464} \\
            w/o sarulab-data     &   \underline{0.484}&              0.750 &              0.718 &              0.516 &                     0.179 &              0.753 &              0.717 & 0.518 &         \underline{0.338} &  \underline{0.755} &  \underline{0.720} &  \underline{0.520} &                        -- &                 -- &                 -- &                 -- \\
            \bottomrule
        \end{tabular}
    }
    \vspace{4mm}
    \subcaption{Utterance-level results at 12\% zoomed-in rate.}
    \scalebox{0.85}{
        \begin{tabular}{l|cccc|cccc|cccc|cccc}
            \toprule
                                 &                                     \multicolumn{4}{c}{BVCC}                      &                                     \multicolumn{4}{|c}{BC}                        &                                           \multicolumn{4}{|c}{SOMOS}                     &                                 \multicolumn{4}{|c}{sarulab-data}                        \\
            Training datasets    &             MSE\zj &             LCC\zk &            SRCC\zk &            KTAU\zk &                    MSE\zj &             LCC\zk &            SRCC\zk &            KTAU\zk &                    MSE\zj &             LCC\zk &            SRCC\zk &            KTAU\zk &                    MSE\zj &             LCC\zk &            SRCC\zk &            KTAU\zk \\
            \midrule
            All datasets         &              0.459 &     \textbf{0.578} &     \textbf{0.579} &     \textbf{0.404} &            \textbf{0.262} &              0.584 &              0.584 &              0.408 &            \textbf{0.234} &              0.579 &              0.579 &              0.403 &            \textbf{0.238} &              0.581 &              0.582 &              0.406 \\
            w/o BVCC             &                 -- &                 -- &                 -- &                 -- &         \underline{0.541} &     \textbf{0.633} &        \textbf{0.626} &  \textbf{0.452} &                     0.324 &     \textbf{0.636} &     \textbf{0.629} &     \textbf{0.454} &                     0.297 &     \textbf{0.636} &      \textbf{0.629}&        \textbf{0.454} \\
            w/o BC         &           \textbf{0.393} &              0.471 &              0.473 &              0.330 &                        -- &                 -- &                 -- &                 -- &                     0.299 &              0.491 &              0.493 &              0.343 &                     0.360 &              0.442 &              0.450 &              0.313 \\
            w/o SOMOS            &              0.447 &  \underline{0.369} &  \underline{0.376} &  \underline{0.257} &                     0.443 &  \underline{0.370} &  \underline{0.375} &  \underline{0.256} &                        -- &                 -- &                 -- &                 -- &         \underline{0.443} &  \underline{0.370} &   \underline{0.378}&        \underline{0.258} \\
            w/o sarulab-data     &  \underline{0.484} &              0.429 &              0.430 &              0.293 &                     0.312 &              0.427 &              0.431 &              0.293 &         \underline{0.392} &  \underline{0.427} &  \underline{0.428} &  \underline{0.292} &                        -- &                 -- &                 -- &                 -- \\
            \bottomrule
        \end{tabular}
    }
    \vspace{4mm}
    \subcaption{System-level results at 12\% zoomed-in rate.}
    \scalebox{0.85}{
        \begin{tabular}{l|cccc|cccc|cccc|cccc}
            \toprule
                                 &              \multicolumn{4}{c}{BVCC}                      &              \multicolumn{4}{|c}{BC}                        &              \multicolumn{4}{|c}{SOMOS}                     &              \multicolumn{4}{|c}{sarulab-data}                        \\
            Training datasets    &             MSE\zj &             LCC\zk &            SRCC\zk &            KTAU\zk &                    MSE\zj &             LCC\zk &            SRCC\zk &            KTAU\zk &                    MSE\zj &             LCC\zk &            SRCC\zk &            KTAU\zk &                    MSE\zj &             LCC\zk &            SRCC\zk &            KTAU\zk \\
            \midrule
            All datasets         &  \underline{0.288} &     \textbf{0.840} &     \textbf{0.854} &      \textbf{0.650} &           \textbf{0.088} &     \textbf{0.844} &     \textbf{0.851} &              0.650 &            \textbf{0.056} &     \textbf{0.840} &              0.844 &              0.642 &            \textbf{0.058} &     \textbf{0.842} &       \textbf{0.838} &              0.634 \\
            w/o BVCC             &                 -- &                 -- &                 -- &                 -- &         \underline{0.343} &              0.823 &              0.832 &     \textbf{0.665} &                     0.128 &              0.824 &     \textbf{0.846} &     \textbf{0.681} &                     0.101 &              0.825 &              0.836 &     \textbf{0.673} \\
            w/o BC         &           \textbf{0.145} &              0.826 &              0.819 &              0.610 &                        -- &                 -- &                 -- &                 -- &                     0.069 &              0.804 &              0.823 &              0.642 &                     0.122 &              0.756 &              0.805 &              0.602 \\
            w/o SOMOS            &              0.224 &  \underline{0.667} &              0.696 &              0.467 &                     0.221 &  \underline{0.665} &              0.682 &      \underline{0.459} &                    -- &                 -- &                 -- &                 -- &         \underline{0.221} &  \underline{0.665} &  \underline{0.700} & \underline{0.483} \\
            w/o sarulab-data     &              0.282 &              0.671 &  \underline{0.647} &      \underline{0.448} &                 0.102 &              0.674 &    \underline{0.661} &    \underline{0.459} &     \underline{0.186} &  \underline{0.675} &  \underline{0.690} &  \underline{0.483} &                        -- &                 -- &                 -- &                 -- \\
            \bottomrule
        \end{tabular}
    }
\end{table*}

%
%
\Section{Conclusion}
\label{sec:conclusion}


In this paper, we presented our automatic MOS prediction system (UTMOSv2) submitted to the VMC 2024.
Our system achieved first place in 7 out of 16 metrics in the VMC 2024 Track 1.
The submitted T05 system leverages the fusion of spectrogram features from a pretrained image feature extractor and speech features from pretrained speech SSL models.
Additionally, multi-stage learning and the use of multiple datasets were introduced.
In the ablation study, we demonstrated that combining spectrogram features and SSL features improves the correlation-based metrics, while the MSE was best when only the spectrogram feature was used.
Furthermore, the use of a wider range of datasets and multi-stage learning enhanced the performance of the MOS prediction.
Future work includes constructing a MOS prediction system not only for the naturalness of synthetic speech but also for other aspects of speech, such as prosody.

\section{Acknowledgements}
This work was supported by JST Moonshot JPMJMS2011. We are also grateful to Kazuki Yamauchi, Osamu Take, and Takaki Hamada of the University of Tokyo for their helpful discussions. 
%

\bibliographystyle{bib/IEEEbib}
\bibliography{bib/refs}

\end{document}